\documentclass[10pt,prb,twocolumn]{revtex4}
\usepackage[english]{babel}
\usepackage{amssymb}
\usepackage{amsfonts}
\usepackage{amsmath}
\usepackage{graphicx}
\usepackage{epsfig}
\newcommand{\be}{\begin{equation}} \newcommand{\ee}{\end{equation}}
\newcommand{\bea}{\begin{eqnarray}} \newcommand{\eea}{\end{eqnarray}}
\newcommand{\beann}{\begin{eqnarray*}}  \newcommand{\eeann}{\end{eqnarray*}}
\newcommand{\bfig}{\begin{figure}} \newcommand{\efig}{\end{figure}}
\newcommand{\ba}{\begin{array}} \newcommand{\ea}{\end{array}}
\newcommand{\bcen}{\begin{center}} \newcommand{\ecen}{\end{center}}
\newcommand{\btab}{\begin{tabular}} \newcommand{\etab}{\end{tabular}}

\begin{document}
\title{Surface Plasmons and Topological Insulators}

\author
{
   Andreas Karch}

\affiliation{
$^1$Department of Physics, University of Washington, Seattle, WA 98195-1560, USA\\
 E-mail:\tt{ karch@phys.washington.edu}
}

\begin{abstract}
We study surface plasmons localized on interfaces between topologically trivial and topologically non-trivial time reversal invariant materials in three dimensions. For the interface between a metal and a topological insulator the magnetic polarization of the surface plasmon is rotated out of the plane of the interface; this effect should be experimentally observable by exciting the surface plasmon with polarized light. More interestingly, we argue that the same effect also is realized on the interface between vacuum and a doped topological insulator with non-vanishing bulk carrier density.
\end{abstract}

\maketitle

\section{Introduction}

Surface plasmons are localized excitations traveling along the interface between two materials\cite{ritchie,stern}. Classically they are only possible if one of the two materials has a real but negative permittivity $\epsilon$, while the other material has real and positive permittivity, e.g. can be taken to be vacuum. A typical example of a material with negative $\epsilon$ is an idealized metal described in terms of a gas of free electrons with
\be
\label{drude}
\frac{\epsilon(\omega)}{\epsilon_0} = 1 - \frac{\omega_p^2}{\omega (\omega + i \gamma)}
\ee
where $\omega_p^2= \frac{n_e e^2}{m \epsilon_0}$ is the plasma frequency (with $n_e$ being the number density of electrons and $m$ their effective mass). The $\gamma$ damping term is a small positive quantity setting the DC conductivity of the metal. For the ideal ($\gamma=0$) Drude gas, this frequency dependent $\epsilon$ is negative for $\omega < \omega_p$. The description in terms of a free electron gas is typically a good description of metals for sufficiently large $\omega$.

In this work we study the properties of these surface plasmons when they are localized on a topologically non-trivial interface. In three spatial dimensions a time reversal invariant band insulator is characterized by a $\mathbb{Z}_2$ valued index characterizing the topology of the bandstructure\cite{fukanemele,roy}. The low energy effective description of such a material contains a topological $\theta \vec{E} \cdot \vec{B}$ term\cite{qhz} leading to modified constitutive relations for Maxwell's equations in matter. The parameter $\theta$ is quantized to take values that are integer multiples of $\pi$. By a topologically non-trivial interface we mean an interface across which $\theta$ jumps.

The simplest topologically non-trivial interface that supports surface plasmons is the interface between a topological insulator and a metal. $\epsilon$ is real and positive in the insulator and, for a range of frequencies, can be taken to be real and negative in the metal. We show that in this case the polarization of the surface plasmon experiences a non-trivial rotation due to the jump in $\theta$ across the interface: while for a standard metal/insulator interface the surface plasmon is entirely ``transverse magnetically" (TM) polarized, with both $\vec{B}$ and $\vec{H}$ lying in the plane of the interface (as well as orthogonal to the direction of propagation of the surface wave), in the topologically non-trivial case the magnetic field is rotated out of the plane of the interface by an angle $\nu_p \propto \alpha \Delta \theta$.

A much more interesting realization of a topological interface is possible if the material with non-vanishing $\theta$ itself is displaying a permittivity of the form \eqref{drude}. This may occur if a material exhibiting the topological bandstructure of a topological insulator has, via doping, acquired a finite bulk carrier density. In fact, basically all experimentally realized 3d TI materials, such as BI$_{1-x}$Sb$_x$ alloys\cite{fukaneprb,hasan1}, Bi$_2$Se$_3$ and Bi$_2$Te$_3$\cite{zhangnature,hasan2,zhang2} have a non-vanishing bulk carrier density. This is often seen as a major obstacle to observing the effects predicted by the low energy description of a topological insulator in terms of an effective theory containing the topological $\theta \vec{E} \cdot \vec{B}$ term. It has however recently been argued\cite{bq,bq2} that even in gapless materials one can still meaningfully define topological indices based on the bandstructure; so correspondingly the topological term in the low energy effective action should still survive. These doped topological insulators (or ``topological metals") should be described by a low energy effective action containing the standard quantized $\theta$ associated with the topological bandstructure together with an $\epsilon(\omega)$ appropriate for a conductor, that is for example of the form \eqref{drude}. Latter is accounting for the free charge carriers. As we will show, such a description will lead to a unique modification of surface plasmon properties, most notably a non-vanishing $\nu_p$.

This effective field theory approach is limited to low frequencies. First one has integrated out modes with energy larger than some energy $E_{gap} =\hbar \omega_{gap}$. These high energy modes had a topological non-trivial band structure and so generated a non-trivial $\theta$ term. The remaining dynamical degrees of freedom, the free charge carriers, are taken into account in terms of $\epsilon(\omega)$. The main focus of our work is on the range of frequencies where $\epsilon(\omega)$ is real and negative. In the Drude model we are for example looking at $\gamma \ll \omega <\omega_p$. As long as there is a hierarchy between $\gamma$ and $\omega_{gap}$, that is $\gamma \ll \omega_{gap}$, there exists a range of frequencies for which our effective description is valid for the range of frequencies in which the novel surface phenomena we describe here occur.
Furthermore, the effective field theory description is only valid if the massless surface modes are in fact gapped by an external T-breaking deformation. One simple experimental way to introduce such a deformation is an external magnetic field $B_z$ orthogonal to the interface. Surface plasmons in the presence of external magnetic fields are well understood\cite{chiuquinn,cq2,wallis}. In particular, for an external field orthogonal to the interface one also finds a non-zero $\nu_p$ which vanishes as $B_z$ is taken to zero. By measuring $\nu_p$ as a function of $B_z$ one can find the topological $\nu_p$ as the constant off-set one approaches as $B_z \rightarrow 0$. This offset has equal magnitude but opposite sign depending on whether $B_z$ approaches zero from above or below. The experimental signature for identifying $\nu_p$ is hence in complete parallel to the proposal of Ref. \onlinecite{qhz} for measuring the Faraday and Kerr rotations associated with topologically non-trivial interfaces between two insulators.

The organization of this paper is as follow. In the following section we will review the classical theory of surface plasmons. In section 3 we then work out the corresponding equations for the case of a topologically non-trivial interface. Most importantly, we obtain an expression for $\nu_p$. In section 4 we discuss important generalizations. In particular, we discuss external magnetic fields and discuss the behavior of $\nu_p$ as a function of $B_z$. We also discuss the Kerr effect in the case of a topological non-trivial metal/insulator interface.

\section{Review of Surface Plasmon Properties}

To find a surface plasmon mode we look for a solution to Maxwell's equations that propagates along the interface but is exponentially damped away from the interface. With the interface at $z=0$ we can take the direction of propagation of the wave, without loss of generality, to be the $x$ direction.

There are two polarizations of surface plasmon modes that can in principle be constructed, even though typically only one of them is possible, as we will explain. Let us first look at ``transverse magnetic" (TM) modes. That is we are looking for a solution of the form
\be
\label{TM}
\vec{H}_i =  H_0 \,  \hat{e}_y  \, e^{i (k_x x - \omega t)} e^{i k^z_i z} = H_0 \,  \hat{e}_y  \, e^{i (k_x x - \omega t)} e^{- |\kappa_{i}^z| z} \ee
where the subscript $i$=$>$, $<$ refers to the two regions $z>0$ and $z<0$. We indicated that for the plasmon to be localized we need $k_i^z = i \kappa_i^z$ to be purely imaginary with positive (negative) imaginary part for $z>0$ ($z<0$).
Solving Maxwell's equations with this ansatz yields the standard dispersion relation
\be
\label{dispersion}
k_x^2 + (k^z_i)^2 = k_x^2 - (\kappa^z_i)^2 = \frac{\omega^2}{c_i^2} = \epsilon_i \mu_i \omega^2.
\ee
To completely specify the solution we need to impose the standard interface boundary conditions following from Maxwell's equations, requiring continuity of $H_{\|}$, $E_{\|}$, $D_{\perp}$ and $B_{\perp}$. Continuity of $H_{\|}$ was built into our ansatz. For topologically trivial materials we have the simple constitutive relations
\be \vec{D} = \epsilon \vec{E}, \quad \quad \quad \vec{H} = \frac{\vec{B}}{\mu} \ee
so $\vec{B}$ is aligned with $\vec{H}$ and $\vec{D}$ with $\vec{E}$.
The transverse magnetic mode $B_{\perp}$ vanishes and so continuity is trivial. From Maxwell's equations we finally have
\be
\label{eandb}
\vec{E} = - \hat{k} \times (c \vec{B}), \quad \quad \quad \vec{D} = - \hat{k} \times  \frac{  \vec{H} }{c}
\ee
so (using $\hat{k} = c \vec{k}/{\omega}$)
\be
\label{fulld}
\vec{D}_i =  \frac{(k_i^z H_0) \,  \hat{e}_x  - (k_x H_0) \, \hat{e}_z }{\omega} \, e^{i (k_x x - \omega t)} e^{i k^z_i z}  \ee
As the coefficient of $\hat{e}_z$ is identical on both sides of the interface, continuity of $D_{\perp}$ is automatically satisfied. Last but not least, continuity of $E_{\|}$ demands
\be
\label{spcondition}
\frac{k^z_>}{\epsilon_>} = \frac{k^z_<}{\epsilon_<} .
\ee
To get the desired solution localized at the defect with opposite signs for $k^z_i$, we clearly need one of the $\epsilon_i$ to be negative (assuming they are real). As we will see, a propagating surface plasmon solution (meaning $k_x$ real, $k_z$ purely imaginary) can only be found if we don't just have opposite signs for the two $\epsilon$'s but satisfy the slightly stronger condition that
\be
\epsilon_> + \epsilon_< < 0.
\ee

The analysis for the transverse electric polarization starts with
\be \vec{E}_i =  E_0 \,  \hat{e}_y  \, e^{i (k_x x - \omega t)}  e^{- |\kappa_{i}^z| z} \ee
and proceeds in parallel to above. One finds that this time $\mu$ will have to take opposite signs on the two sides of the interface for this mode to exist. This is much harder to achieve in practice (even though for certain frequencies it can be done in meta-materials). For the rest of this work we are going to assume that all $\mu$ are real and positive and so no transverse electrically polarized surface plasmons exist.

Once the surface plasmon condition \eqref{spcondition} is met, one can solve for the properties of the surface plasmon using \eqref{spcondition} together with the dispersion relation \eqref{dispersion}:
\be k_x = \omega \sqrt{\frac{\epsilon_> \epsilon_< (\epsilon_> \mu_< - \epsilon_< \mu_>)}{(\epsilon_> + \epsilon_<)
(\epsilon_> - \epsilon_<)}} \approx \omega \sqrt{\mu \frac{\epsilon_> \epsilon_< }{(\epsilon_> + \epsilon_<)}} \ee
where in the last step we specialized to the case $\mu_> \approx \mu_< = \mu$. Last but not least (for equal $\mu$)
\be
\kappa_i= \sqrt{\frac{ - \mu \epsilon_i^2}{\epsilon_> + \epsilon_<}}.
\ee
We see that, as we advertised above, propagating surface plasmons exist for real $\epsilon_i$ when $\epsilon_> \epsilon_< <0$ (the two have opposite signs) and in addition $\epsilon_> + \epsilon_< <0 $ (the negative $\epsilon$ is larger in magnitude).

\begin{figure}[h]
\begin{center}
\includegraphics[scale=0.9]{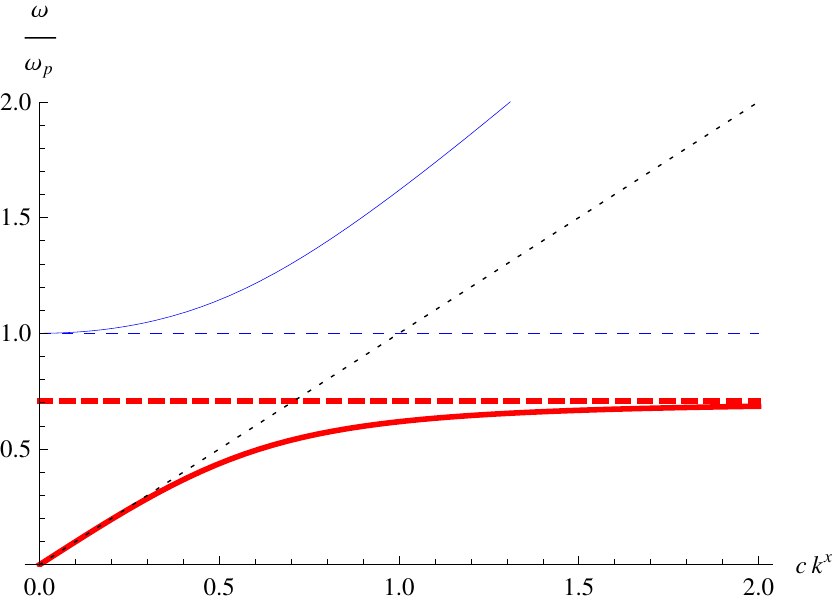}
\end{center}
\caption{
\label{fig1} Dispersion relation for the surface plasmon (thick solid red line) and the bulk plasmon (thin solid blue) for a planar interface between an ideal Drude metal ($\gamma=0$) and vacuum. Also indicated are the lines $\omega=ck^x$ (black dotted line), $\omega=\omega_p$ (thin dashed blue line) and $\omega=\omega_p/\sqrt{2}$ (thick dashed red line).
}
\end{figure}

As an example, take an interface between vacuum ($\epsilon_> = \epsilon_0$, $\mu_>=\mu_0$) and a simple ideal Drude metal, that is $\mu_<=\mu_0$ and $\epsilon_<$ being given by the Drude relation \eqref{drude} with $\gamma=0$. The resulting surface plasmon dispersion relation is
\begin{eqnarray}
\nonumber
k_x(\omega) &=& \frac{\omega}{c} \sqrt{\frac{\omega^2 - \omega_p^2}{2 \omega^2 - \omega_p^2}}, \\
\omega^2(k_x) &=& \frac{\omega_p^2}{2} + c^2 k_x^2 - \sqrt{\omega_p^4/4 + c^4 k_x^4}.
\end{eqnarray}
This solution describes the propagating surface plasmon which exists for $0 < \omega < \omega_p/\sqrt{2}$.
There is a second propagating solution for $\omega^2$ with a + sign in front of the root that exists for $\omega > \omega_p$. This is simply a propagating bulk wave in the Drude metal, which for $\omega>\omega_p$ has a positive $\epsilon$. No propagating solutions exist in the
window $\omega_p/\sqrt{2} < \omega < \omega_p$. The whole spectrum is displayed in Figure \ref{fig1}. For small non-zero $\gamma$ the very low frequency region with $\omega \sim \gamma$ experiences some damping. For sufficiently small $\gamma$ the ideal Drude analysis of the surface plasmon applies in the regime $\gamma \ll \omega \ll \omega_p/\sqrt{2}$. For very large $k_x$ (of order the Fermi momentum $k_f$ of the free charge carriers), where $\omega \rightarrow \omega_p/\sqrt{2}$, the simple model also breaks down. We can no longer treat the metal in terms of a local permittivity $\epsilon$ and need to take into account non-local effects.

\section{Surface Plasmons in Topological Insulators}

In SI units the constitutive relations for a topological insulator read
 \begin{eqnarray}
 \nonumber
 \vec{D} &=& \epsilon \vec{E} -  \epsilon_0 \alpha \frac{\theta}{\pi} (c_0 \vec{B}) \\ \label{SITI}
 c_0 \vec{H} &=& \frac{ c_0 \vec{B}}{\mu}  +  \alpha \frac{\theta}{\pi} \frac{ \vec{E}}{\mu_0}
\end{eqnarray}
with $c_0^{-2}=\mu_0 \epsilon_0$ being the vacuum speed of light.
The description of a TI in terms of the modified constitutive relations incorporating the topological magneto-electric effect is only valid when the massless topological surface modes are gapped. One option for a time-reversal breaking deformation that gaps the surface modes is an external magnetic field in the $z$ direction. In this section we will first demonstrate the effect of the modified constitutive relations without explicitly including an external magnetic field. As for Faraday and Kerr rotation, measuring the surface plasmon dispersion relation as a function of external magnetic field is an experimentally easy to define way to pin-point the topological properties of the material.

There are two configurations we are interested in: (A) the interface between a genuine topological insulator and a metal and (B) the interface between vacuum and a topological ``insulator" with a non-vanishing bulk conductivity. For (A) we have a positive $\epsilon$ inside the TI while $\epsilon$ in the metal can be negative, supporting the surface plasmon. For (B) we have $\epsilon=\epsilon_0$ positive for vacuum, but $\epsilon_<$ inside the conducting TI can be negative, again supporting a surface plasmon. Doing the analysis for general $\epsilon_>$ and $\epsilon_<$ and $\Delta \theta = \theta_< - \theta_>$ automatically includes both cases.

We start with the most general wave ansatz for $\vec{E}$, allowing both for TE and TM polarizations; it is easy to see that a ansatz that is purely TM or purely TE on both sides of the interface is too restrictive to allow solutions:
\be
\label{eansatz}
\vec{E}_i =
\left [ E_0 \, \hat{e}_y + \frac{c_i E^i_1}{\omega} \, (k_x \, \hat{e}_z - k_i^z \, \hat{e}_x) \right ] \, e^{i (k_x x - \omega t)} e^{i k^z_i z}
\ee
Once more we are interested in a solution where $k^z_i$ is purely imaginary with opposite signs on the two sides of the interface. Our ansatz builds in continuity of $E_{y}$, continuity of $E_{x}$ gives
us a non-trivial condition on the TM polarization:
\be
\label{TMcondition}
c_> E^>_1 k_>^z = c_< E^<_1 k_<^z.
\ee
Maxwell's equations themselves are unmodified (only the constitutive relations were changed). Furthermore, we still have plane wave solutions with the dispersion relation \eqref{dispersion} and the relations between $\vec{E}$ and $\vec{B}$ ($\vec{D}$ and $\vec{H}$) of \eqref{eandb}.
So we get in analogy with \eqref{fulld}
\be
\label{dTI}
\vec{B}_i =
\left [ \frac{   (k_x E_0) \, \hat{e}_z -(k_i^z E_0) \, \hat{e}_x }{\omega} - \frac{E_1^i}{c_i} \, \hat{e}_y \right ] \, e^{i (k_x x - \omega t)} e^{i k^z_i z}
\ee
where we had to use the dispersion relation \eqref{dispersion}.
Continuity of $B_{\perp}=B_z$ is automatically satisfied.
$\vec{D}$ and $\vec{H}$ follow from the constitutive relation \eqref{SITI}.
Continuity of $D_{\perp}$ requires
\be
\label{dzcondition}
\epsilon_> c_> E^>_1 - \epsilon_0 \alpha \frac{\theta_>}{\pi} (c_0 E_0) =
\epsilon_< c_< E^<_1 - \epsilon_0 \alpha \frac{\theta_<}{\pi} (c_0  E_0)
\ee
which, using our earlier relation \eqref{TMcondition} can be used to solve for the TE to TM amplitude ratio ($\Delta \theta = \theta_<-\theta_>$) :
\be
\label{TETM}
E^<_1 = \frac{ \alpha (\Delta \theta/\pi) \epsilon_0 c_0}{
c_< \left [ \epsilon_<  - \epsilon_> (k_<^z/k_>^z) \right ]}
E_0.
\ee
Last but not least, continuity of $H_{\|}$ gives us two more conditions, one for the $x$
\be
\label{excondition}
k^z_> \left ( \frac{E_0}{ \mu_>} + \alpha \frac{\theta_>}{\pi} \frac{c_> E_1^>}{\mu_0 c_0} \right ) =
k^z_< \left ( \frac{E_0}{ \mu_<} + \alpha \frac{\theta_<}{\pi} \frac{c_< E_1^<}{\mu_0 c_0}\right )
\ee
and one for the $y$ component
\be
\label{eycondition}
-\frac{E_1^>}{c_> \mu_>} + \alpha \frac{\theta_>}{\pi} \frac{E_0}{\mu_0 c_0}
=
-\frac{E_1^<}{c_< \mu_<} + \alpha \frac{\theta_<}{\pi} \frac{E_0}{\mu_0 c_0}.
\ee
Reassuringly the condition on $H_y$ is redundant with what we found above in \eqref{TETM} from the continuity of $D_{\perp}$. The $H_x$ condition can be solved by a relation similar to \eqref{TETM} for the TE to TM amplitude ratio:
\be
\label{TETMtwo}
E^<_1 = \frac{\mu_< (k^z_>/k^z_<) - \mu_>}{ \alpha (\Delta \theta/\pi)  \mu_> \mu_< c_<}
\mu_0 c_0 \, E_0.
\ee
For the two equations relating TE and TM amplitude, \eqref{TETM} and \eqref{TETMtwo} to be simultaneously satisfied we need
\be
\left ( \frac{ \alpha \Delta \theta}{\pi} \right )^2 \frac{\epsilon_0 \mu_> \mu_<}{\mu_0}
= \left [ \epsilon_< - \epsilon_> (k^z_</k^z_>) \right ] \left [ \mu_< (k^z_>/k^z_<) - \mu_>
\right ].
\ee
For $\Delta \theta=0$ this nicely yields two polarizations, the standard equations for our TM polarized surface modes \eqref{spcondition} as well as the magnetic analog with TE polarization and
\be
\label{magenticspcondition}
\mu_< k^z_> = \mu_> k^z_<.
\ee
As mentioned before, latter would require negative $\mu$ and so typically this TE surface plasmon is not realized. Including the effects of $\Delta \theta$ the two solutions become:
\begin{eqnarray}
\label{TIplasmonfull}
x&&\equiv \frac{k^z_<}{k^z_>} = \frac{1}{2} \left [
\left ( \frac{\alpha \Delta \theta}{\pi} \right )^2 \frac{\epsilon_0  \mu_<}{ \epsilon_> \mu_0}
+ \frac{\epsilon_<}{\epsilon_>} + \frac{\mu_<}{\mu_>} \right . \\
&& \left . \pm \sqrt{
-4\frac{\epsilon_< \mu_<}{\epsilon_> \mu_>} + \left ( \left ( \frac{\alpha \Delta \theta}{\pi} \right )^2 \frac{\epsilon_0 \mu_<}{\epsilon_> \mu_0} + \left (
\frac{\epsilon_<}{\epsilon_>} + \frac{\mu_<}{\mu_>}
\right) \right )^2
}
\right ].
\nonumber
\end{eqnarray}
For $\Delta \theta=0$ this correctly reduces to $k^z_</k^z_> = \epsilon_</\epsilon_>$ and
$k^z_</k^z_> = \mu_</\mu_>$ respectively. For materials with no significant magnetism ($\mu \sim \mu_0$) the latter does not give rise to a propagating surface plasmon. Specializing to $\mu_> = \mu_< = \mu_0$, \eqref{TIplasmonfull} simplifies to
\begin{eqnarray}
\label{TIplasmon}
x&&\equiv \frac{k^z_<}{k^z_>} = \frac{1}{2} \left [
\left ( \frac{\alpha \Delta \theta}{\pi} \right )^2 \frac{\epsilon_0 }{ \epsilon_> }
+ \frac{\epsilon_<}{\epsilon_>} + 1 \right . \\
&& \left . \pm \sqrt{
-4\frac{\epsilon_< }{\epsilon_>} + \left ( \left ( \frac{\alpha \Delta \theta}{\pi} \right )^2 \frac{\epsilon_0 }{\epsilon_> } + \left (
\frac{\epsilon_<}{\epsilon_>} + 1
\right) \right )^2
}
\right ].
\nonumber
\end{eqnarray}
To see how this affects the dispersion relation note that for any surface plasmon like solution
with $\mu_> = \mu_< = \mu_0$ we get directly from the dispersion relations \eqref{dispersion}
\be
\label{gendispersion}
k_x = \omega \sqrt{\mu_0 \frac{\epsilon_< - x^2 \epsilon_>}{1- x^2}}
\ee
where for us $x$ will be given by the negative sign solution of \eqref{TIplasmon}. Taking $\alpha=1/137$, $\Delta \theta = \pi$ and one $\epsilon$ to be $\epsilon_0$, while the other is given by the ideal ($\gamma=0$) Drude formula \eqref{drude} leads to a dispersion spectrum that is almost indistinguishable from the one plotted in Figure \ref{fig1} for the topological trivial case (as $\alpha \pi \ll 1$). One can however notice some quantitative differences. One interesting effect is that the asymptotic (large $k_x$) frequency of the surface plasmon shifts from its Drude value of $\omega_{\infty} = \omega_p/\sqrt{2}$ to
\be
\label{omshift}
\omega_{\infty} = \frac{\omega_P}{\sqrt{2}} \frac{1}{\sqrt{1+ \left ( \frac{\alpha \Delta \theta}{2\pi} \right )^2}}.
\ee
The most notable and probably experimentally most easily accessible feature is that the polarization of the surface plasma wave is altered. Unlike the ${\cal O}(\alpha^2)$ modifications to the dispersion relation this change in polarization occurs already at ${\cal O}(\alpha)$. Furthermore, it is sensitive to the sign of $\Delta \theta$ and so it will change direction as the external T-breaking deformation that gapped the massless fermionic surface modes changes sign. While in the topological trivial case the surface plasmon wave is completely polarized in the TM direction and correspondingly $\vec{E} \cdot \hat{e}_y \sim E_0 =0$, in the topological non-trivial wave the surface plasmon acquires a non-vanishing TE component given by \eqref{TETM}:
\begin{eqnarray}
\label{nup}
\tan(\nu_p) &\equiv& \frac{E_0}{E_1^<} = \frac{c_< \left ( \epsilon_<  - \epsilon_> x \right )}{\alpha (\Delta \theta/\pi) c_0 \epsilon_0} \\ &=& \frac{\alpha \Delta \theta}{\pi} \frac{c_< \epsilon_<}{c_0 (\epsilon_>-\epsilon_<)}  + {\cal O} \left [ \left (\frac{\alpha \Delta \theta}{\pi} \right )^2 \right ] .
\nonumber
\end{eqnarray}
This angle should be measurable experimentally by trying to couple polarized light into surface plasmons. Surface plasmons can not directly be excited by light; the plasmon dispersion curve always lies {\it below} the dispersion of light, so there is no direct coupling of light into plasmons. But it is well known how to circumvent this problem: coupling the light via a grating\cite{grating} or, equivalently, through a rough surface allows Umklapp processes that can change the momentum of the photon to couple to the surface plasmon.
Alternatively, ``attenuated total reflection" can be employed\cite{otto,kretschmann} to provide the incoming radiation with an imaginary wave vector in the direction perpendicular to the surface. In either case, tuning the polarization of the incoming radiation one should find a maximum in the light-plasmon coupling when the polarization of the external light has maximal overlap with the polarization of the allowed plasmon modes as characterized by $\nu_p$.

\section{Generalizations: External Fields and Kerr Effect}
\subsection{External Magnetic Fields}

The effective field theory description in terms of the modified constitutive relations \eqref{SITI} is only valid if the surface modes are gapped by an external, T-breaking perturbation. In particular, whether $\theta$ should be taken to be $+\pi$ or $-\pi$ depends on the sign on the external perturbation. Note that while the modification of the plasmon dispersion relation only depends on $(\Delta \theta)^2$ and so is insensitive to the sign of $\theta$, the plasmon polarization angle $\nu_p$ of \eqref{nup} actually changes sign as $\theta$ changes from $\pi$ to $-\pi$ and so is sensitive to the external T-breaking perturbation.

A simple way to gap the surface modes is to turn on an external magnetic field perpendicular to the interface, $\vec{B}_{ext} = B_z \hat{e}_z$ in our case. The sign of $B_z$ determines the sign of $\theta$ and hence the sign of $\nu_p$. An external magnetic field will also give rise to a non-trivial $\nu_p$ that grows with the magnetic
field\cite{chiuquinn,cq2,wallis}. The effect of the orthogonal magnetic field is to modify the constitutive relations, as $\epsilon$ depends on $B_z$. Most notably, magnetic fields force us to treat $\epsilon$ as a tensor and, in particular, give rise to a non-trivial $\epsilon_{xy}$. For the ideal Drude metal one has for example
\be
\epsilon= \epsilon_0 \left ( \begin{array}{ccc}
1 + \frac{\omega_p^2}{\omega_c^2-\omega^2} & i \frac{\omega_c \omega_p^2}{\omega (\omega^2-\omega_c^2)} & 0\\
 -i \frac{\omega_c \omega_p^2}{\omega (\omega^2-\omega_c^2)}
 & 1 + \frac{\omega_p^2}{\omega_c^2-\omega^2} & 0\\
 0& 0&1-\frac{\omega_p^2}{\omega^2}  \\
\end{array} \right )
\ee
The resulting linear equations for the surface plasmon turn out to be rather cumbersome, but the general solution has been presented quite comprehensively in Ref. \onlinecite{wallis}. The details of the analysis are not important for us, but the crucial point is that $\nu_p$ is non-zero at finite $B_z$. While it in general depends on $B_z$ in a non-linear fashion, it does nicely go to zero as $B_z$ goes to zero.

So in order to experimentally measure the topological properties of the plasmon we can follow the same strategy as proposed in Ref. \onlinecite{qhz}: one needs to measure $\nu_p$ as a function of external magnetic field. For topologically trivial interfaces ($\Delta \theta=0$) $\nu_p$ will go to zero linearly with $B_z$ for small $B_z$, whereas for a topologically non-trivial interface it will tend to a constant. This constant has opposite sign depending on whether we dial $B_z$ to zero from $B_z>0$ or from $B_z<0$. Around $B_z \sim 0$ the polarization experiences a rapid transition from $\nu_p$ given by \eqref{nup} to $-\nu_p$.

\subsection{Kerr Effect}

For constant (that is frequency independent) $\epsilon$ and $\mu$ it was shown\cite{qhz,Karch:2009sy} that light reflecting off a topological non-trivial interface experiences a non-trivial Kerr rotation of its polarization. In addition, the light transmitted through the interface also experienced a non-trivial Faraday rotation. A suitable linear combination of the two angles has been
shown\cite{joseph,zhangalpha} to be quantized in units of $\alpha$. This quantization is robust even after the effects of the finite thickness of the film are taken into account. In our case, where the frequency dependent $\epsilon$ in the frequency range of interest ($\omega<\omega_p$) is negative, there is clearly no transmitted wave (the only propagating degrees of freedom with $\omega<\omega_p$ are the surface plasmons). Light with $\omega < \omega_p$ is completely reflected (to couple it into surface plasmons one needs to utilize a grating, as we mentioned before. Without the grating, light is completely reflected). This reflected light of course still experiences a Kerr rotation. The angle of the Kerr rotation $\theta_K$ is still given by the identical expression as in \onlinecite{qhz,Karch:2009sy}. For a light in-coming at orthogonal incidence from (say) $z>0$ we have (in SI units):
\be
\tan(\theta_K) = Y_>  \frac{ 2 \frac{\alpha \Delta \theta}{\pi} Y_0}
{Y_<^2 - Y_>^2 + \left ( \frac{\alpha \Delta \theta}{\pi} \right )^2 Y_0^2 }
\ee
where $Y_i \equiv \sqrt{\epsilon_i/\mu_i}$ is the admittance of the two media and vacuum respectively.
As long as $\epsilon_>$ and $\mu_>$ are real and positive, we can send in such an incoming wave. For negative $\epsilon_<$ we have $Y_<^2$ real and negative as well, but this still yields a perfectly real and well defined $\theta_k$. The main difference to the case of a wave reflecting of a dielectric is that wavevector $\vec{k} = k_z \hat{e}_z$ is purely imaginary for $z<0$ and there is no transmitted wave. This does not affect the calculation of the Kerr angle. Kerr effect and the polarization of surface plasmons however are clear experimental signatures of a topological metal.

For $\omega > \omega_p$ the ideal Drude metal behaves like a dielectric with both $\mu$ and $\epsilon$ real and positive. In this case, transmitted and reflected wave do exist and light experiences the standard Kerr and Faraday rotations associated with topological insulators. Unless the material has a built in hierarchy with $\omega_{gap} \gg \omega_p$, this high frequency regime of the Drude metal is however beyond the validity of our effective field theory.

\section*{Acknowledgements}
Thanks to Keiko Torii for introducing me to the concept of surface plasmons and to Jerry Seidler for discussions about experimental realizations. Special thanks to Zohreh Davoudi for correcting a mistake in equation \eqref{omshift} in an earlier version of this manuscript. The work of AK was supported in part by the U.S. Department of Energy under Grant No.~DE-FG02-96ER40956.

\bibliography{plasmon}

\begin{thebibliography}{21}
\expandafter\ifx\csname natexlab\endcsname\relax\def\natexlab#1{#1}\fi
\expandafter\ifx\csname bibnamefont\endcsname\relax
  \def\bibnamefont#1{#1}\fi
\expandafter\ifx\csname bibfnamefont\endcsname\relax
  \def\bibfnamefont#1{#1}\fi
\expandafter\ifx\csname citenamefont\endcsname\relax
  \def\citenamefont#1{#1}\fi
\expandafter\ifx\csname url\endcsname\relax
  \def\url#1{\texttt{#1}}\fi
\expandafter\ifx\csname urlprefix\endcsname\relax\def\urlprefix{URL }\fi
\providecommand{\bibinfo}[2]{#2}
\providecommand{\eprint}[2][]{\url{#2}}

\bibitem[{\citenamefont{Ritchie}(1957)}]{ritchie}
\bibinfo{author}{\bibfnamefont{R.~H.} \bibnamefont{Ritchie}},
  \bibinfo{journal}{Phys. Rev.} \textbf{\bibinfo{volume}{106}},
  \bibinfo{pages}{874} (\bibinfo{year}{1957}).

\bibitem[{\citenamefont{Stern and Ferrell}(1960)}]{stern}
\bibinfo{author}{\bibfnamefont{E.~A.} \bibnamefont{Stern}} \bibnamefont{and}
  \bibinfo{author}{\bibfnamefont{R.~A.} \bibnamefont{Ferrell}},
  \bibinfo{journal}{Phys. Rev.} \textbf{\bibinfo{volume}{120}},
  \bibinfo{pages}{130} (\bibinfo{year}{1960}).

\bibitem[{\citenamefont{Fu et~al.}(2007)\citenamefont{Fu, Kane, and
  Mele}}]{fukanemele}
\bibinfo{author}{\bibfnamefont{L.}~\bibnamefont{Fu}},
  \bibinfo{author}{\bibfnamefont{C.~L.} \bibnamefont{Kane}}, \bibnamefont{and}
  \bibinfo{author}{\bibfnamefont{E.~J.} \bibnamefont{Mele}},
  \bibinfo{journal}{Phys. Rev. Lett.} \textbf{\bibinfo{volume}{98}},
  \bibinfo{pages}{106803} (\bibinfo{year}{2007}).

\bibitem[{\citenamefont{Roy}(2009)}]{roy}
\bibinfo{author}{\bibfnamefont{R.}~\bibnamefont{Roy}}, \bibinfo{journal}{Phys.
  Rev. B} \textbf{\bibinfo{volume}{79}}, \bibinfo{pages}{195322}
  (\bibinfo{year}{2009}).

\bibitem[{\citenamefont{{Qi} et~al.}(2008)\citenamefont{{Qi}, {Hughes}, and
  {Zhang}}}]{qhz}
\bibinfo{author}{\bibfnamefont{X.}~\bibnamefont{{Qi}}},
  \bibinfo{author}{\bibfnamefont{T.~L.} \bibnamefont{{Hughes}}},
  \bibnamefont{and} \bibinfo{author}{\bibfnamefont{S.}~\bibnamefont{{Zhang}}},
  \bibinfo{journal}{\prb} \textbf{\bibinfo{volume}{78}},
  \bibinfo{pages}{195424} (\bibinfo{year}{2008}), \eprint{0802.3537}.

\bibitem[{\citenamefont{Fu and Kane}(2007)}]{fukaneprb}
\bibinfo{author}{\bibfnamefont{L.}~\bibnamefont{Fu}} \bibnamefont{and}
  \bibinfo{author}{\bibfnamefont{C.~L.} \bibnamefont{Kane}},
  \bibinfo{journal}{Phys. Rev. B} \textbf{\bibinfo{volume}{76}},
  \bibinfo{pages}{045302} (\bibinfo{year}{2007}).

\bibitem[{\citenamefont{{Hsieh} et~al.}(2008)\citenamefont{{Hsieh}, {Qian},
  {Wray}, {Xia}, {Hor}, {Cava}, and {Hasan}}}]{hasan1}
\bibinfo{author}{\bibfnamefont{D.}~\bibnamefont{{Hsieh}}},
  \bibinfo{author}{\bibfnamefont{D.}~\bibnamefont{{Qian}}},
  \bibinfo{author}{\bibfnamefont{L.}~\bibnamefont{{Wray}}},
  \bibinfo{author}{\bibfnamefont{Y.}~\bibnamefont{{Xia}}},
  \bibinfo{author}{\bibfnamefont{Y.~S.} \bibnamefont{{Hor}}},
  \bibinfo{author}{\bibfnamefont{R.~J.} \bibnamefont{{Cava}}},
  \bibnamefont{and} \bibinfo{author}{\bibfnamefont{M.~Z.}
  \bibnamefont{{Hasan}}}, \bibinfo{journal}{\nat}
  \textbf{\bibinfo{volume}{452}}, \bibinfo{pages}{970} (\bibinfo{year}{2008}),
  \eprint{0902.1356}.

\bibitem[{\citenamefont{{Zhang} et~al.}(2009)\citenamefont{{Zhang}, {Liu},
  {Qi}, {Dai}, {Fang}, and {Zhang}}}]{zhangnature}
\bibinfo{author}{\bibfnamefont{H.}~\bibnamefont{{Zhang}}},
  \bibinfo{author}{\bibfnamefont{C.-X.} \bibnamefont{{Liu}}},
  \bibinfo{author}{\bibfnamefont{X.-L.} \bibnamefont{{Qi}}},
  \bibinfo{author}{\bibfnamefont{X.}~\bibnamefont{{Dai}}},
  \bibinfo{author}{\bibfnamefont{Z.}~\bibnamefont{{Fang}}}, \bibnamefont{and}
  \bibinfo{author}{\bibfnamefont{S.-C.} \bibnamefont{{Zhang}}},
  \bibinfo{journal}{Nature Phys.} \textbf{\bibinfo{volume}{5}},
  \bibinfo{pages}{438} (\bibinfo{year}{2009}).

\bibitem[{\citenamefont{{Xia} et~al.}(2009)\citenamefont{{Xia}, {Qian},
  {Hsieh}, {Wray}, {Pal}, {Lin}, {Bansil}, {Grauer}, {Hor}, {Cava}
  et~al.}}]{hasan2}
\bibinfo{author}{\bibfnamefont{Y.}~\bibnamefont{{Xia}}},
  \bibinfo{author}{\bibfnamefont{D.}~\bibnamefont{{Qian}}},
  \bibinfo{author}{\bibfnamefont{D.}~\bibnamefont{{Hsieh}}},
  \bibinfo{author}{\bibfnamefont{L.}~\bibnamefont{{Wray}}},
  \bibinfo{author}{\bibfnamefont{A.}~\bibnamefont{{Pal}}},
  \bibinfo{author}{\bibfnamefont{H.}~\bibnamefont{{Lin}}},
  \bibinfo{author}{\bibfnamefont{A.}~\bibnamefont{{Bansil}}},
  \bibinfo{author}{\bibfnamefont{D.}~\bibnamefont{{Grauer}}},
  \bibinfo{author}{\bibfnamefont{Y.~S.} \bibnamefont{{Hor}}},
  \bibinfo{author}{\bibfnamefont{R.~J.} \bibnamefont{{Cava}}},
  \bibnamefont{et~al.}, \bibinfo{journal}{Nature Physics}
  \textbf{\bibinfo{volume}{5}}, \bibinfo{pages}{398} (\bibinfo{year}{2009}),
  \eprint{0908.3513}.

\bibitem[{\citenamefont{Chen et~al.}(2009)\citenamefont{Chen, Analytis, Chu,
  Liu, Mo, Qi, Zhang, Lu, Dai, Fang et~al.}}]{zhang2}
\bibinfo{author}{\bibfnamefont{Y.~L.} \bibnamefont{Chen}},
  \bibinfo{author}{\bibfnamefont{J.~G.} \bibnamefont{Analytis}},
  \bibinfo{author}{\bibfnamefont{J.-H.} \bibnamefont{Chu}},
  \bibinfo{author}{\bibfnamefont{Z.~K.} \bibnamefont{Liu}},
  \bibinfo{author}{\bibfnamefont{S.-K.} \bibnamefont{Mo}},
  \bibinfo{author}{\bibfnamefont{X.~L.} \bibnamefont{Qi}},
  \bibinfo{author}{\bibfnamefont{H.~J.} \bibnamefont{Zhang}},
  \bibinfo{author}{\bibfnamefont{D.~H.} \bibnamefont{Lu}},
  \bibinfo{author}{\bibfnamefont{X.}~\bibnamefont{Dai}},
  \bibinfo{author}{\bibfnamefont{Z.}~\bibnamefont{Fang}}, \bibnamefont{et~al.},
  \bibinfo{journal}{Science} \textbf{\bibinfo{volume}{325}},
  \bibinfo{pages}{178} (\bibinfo{year}{2009}).

\bibitem[{\citenamefont{{Barkeshli} and {Qi}}(2011)}]{bq}
\bibinfo{author}{\bibfnamefont{M.}~\bibnamefont{{Barkeshli}}} \bibnamefont{and}
  \bibinfo{author}{\bibfnamefont{X.}~\bibnamefont{{Qi}}},
  \bibinfo{journal}{ArXiv e-prints}  (\bibinfo{year}{2011}),
  \eprint{1101.3104}.

\bibitem[{\citenamefont{{Bergman}}(2011)}]{bq2}
\bibinfo{author}{\bibfnamefont{D.~L.} \bibnamefont{{Bergman}}},
  \bibinfo{journal}{ArXiv e-prints}  (\bibinfo{year}{2011}),
  \eprint{1101.4233}.

\bibitem[{\citenamefont{Chiu and Quinn}(1972)}]{chiuquinn}
\bibinfo{author}{\bibfnamefont{K.}~\bibnamefont{Chiu}} \bibnamefont{and}
  \bibinfo{author}{\bibfnamefont{J.}~\bibnamefont{Quinn}}, \bibinfo{journal}{Il
  Nuovo Cimento B (1971-1996)} \textbf{\bibinfo{volume}{10}},
  \bibinfo{pages}{1} (\bibinfo{year}{1972}).

\bibitem[{\citenamefont{Chiu and Quinn}(1974)}]{cq2}
\bibinfo{author}{\bibfnamefont{K.~W.} \bibnamefont{Chiu}} \bibnamefont{and}
  \bibinfo{author}{\bibfnamefont{J.~J.} \bibnamefont{Quinn}},
  \bibinfo{journal}{Phys. Rev. B} \textbf{\bibinfo{volume}{9}},
  \bibinfo{pages}{4724} (\bibinfo{year}{1974}).

\bibitem[{\citenamefont{Wallis et~al.}(1974)\citenamefont{Wallis, Brion,
  Burstein, and Hartstein}}]{wallis}
\bibinfo{author}{\bibfnamefont{R.~F.} \bibnamefont{Wallis}},
  \bibinfo{author}{\bibfnamefont{J.~J.} \bibnamefont{Brion}},
  \bibinfo{author}{\bibfnamefont{E.}~\bibnamefont{Burstein}}, \bibnamefont{and}
  \bibinfo{author}{\bibfnamefont{A.}~\bibnamefont{Hartstein}},
  \bibinfo{journal}{Phys. Rev. B} \textbf{\bibinfo{volume}{9}},
  \bibinfo{pages}{3424} (\bibinfo{year}{1974}).

\bibitem[{\citenamefont{Teng and Stern}(1967)}]{grating}
\bibinfo{author}{\bibfnamefont{Y.-Y.} \bibnamefont{Teng}} \bibnamefont{and}
  \bibinfo{author}{\bibfnamefont{E.~A.} \bibnamefont{Stern}},
  \bibinfo{journal}{Phys. Rev. Lett.} \textbf{\bibinfo{volume}{19}},
  \bibinfo{pages}{511} (\bibinfo{year}{1967}).

\bibitem[{\citenamefont{Otto}(1968)}]{otto}
\bibinfo{author}{\bibfnamefont{A.}~\bibnamefont{Otto}},
  \bibinfo{journal}{Zeitschrift für Physik A Hadrons and Nuclei}
  \textbf{\bibinfo{volume}{216}}, \bibinfo{pages}{398} (\bibinfo{year}{1968}),
  ISSN \bibinfo{issn}{0939-7922}.

\bibitem[{\citenamefont{Kretschmann and Raether}(1968)}]{kretschmann}
\bibinfo{author}{\bibfnamefont{E.}~\bibnamefont{Kretschmann}} \bibnamefont{and}
  \bibinfo{author}{\bibfnamefont{H.}~\bibnamefont{Raether}},
  \bibinfo{journal}{Zeitschrift {f\"ur} Naturforschung A}
  \textbf{\bibinfo{volume}{23}}, \bibinfo{pages}{2135} (\bibinfo{year}{1968}).

\bibitem[{\citenamefont{Karch}(2009)}]{Karch:2009sy}
\bibinfo{author}{\bibfnamefont{A.}~\bibnamefont{Karch}},
  \bibinfo{journal}{Phys.Rev.Lett.} \textbf{\bibinfo{volume}{103}},
  \bibinfo{pages}{171601} (\bibinfo{year}{2009}), \eprint{0907.1528}.

\bibitem[{\citenamefont{{Maciejko} et~al.}(2010)\citenamefont{{Maciejko}, {Qi},
  {Drew}, and {Zhang}}}]{joseph}
\bibinfo{author}{\bibfnamefont{J.}~\bibnamefont{{Maciejko}}},
  \bibinfo{author}{\bibfnamefont{X.}~\bibnamefont{{Qi}}},
  \bibinfo{author}{\bibfnamefont{H.~D.} \bibnamefont{{Drew}}},
  \bibnamefont{and} \bibinfo{author}{\bibfnamefont{S.}~\bibnamefont{{Zhang}}},
  \bibinfo{journal}{Physical Review Letters} \textbf{\bibinfo{volume}{105}},
  \bibinfo{pages}{166803} (\bibinfo{year}{2010}), \eprint{1004.2514}.

\bibitem[{\citenamefont{{Lan} et~al.}(2011)\citenamefont{{Lan}, {Wan}, and
  {Zhang}}}]{zhangalpha}
\bibinfo{author}{\bibfnamefont{Y.}~\bibnamefont{{Lan}}},
  \bibinfo{author}{\bibfnamefont{S.}~\bibnamefont{{Wan}}}, \bibnamefont{and}
  \bibinfo{author}{\bibfnamefont{S.}~\bibnamefont{{Zhang}}},
  \bibinfo{journal}{ArXiv e-prints}  (\bibinfo{year}{2011}),
  \eprint{1101.0314}.

\end{thebibliography}

\end{document}